\documentstyle[aps,preprint,tighten]{revtex}

\include{epsf}
\newcommand{\eq}{\begin{equation}}
\newcommand{\en}{\end{equation}}
\newcommand{\eqa}{\begin{eqnarray}}
\newcommand{\ena}{\end{eqnarray}}
\newcommand{\ml}{\begin{mathletters}}
\newcommand{\eml}{\end{mathletters}}

\begin{document}

\draft

\title{Stability of the replica symmetric solution for the information
conveyed by a neural network}

\author{Simon Schultz\dag\ and Alessandro Treves\ddag}

\address{\dag\ Department of Experimental Psychology, South Parks Rd.,
University of Oxford, Oxford OX1 3UD, U.K.}

\address{\ddag\ Programme in Neuroscience, International School for Advanced
Studies, via Beirut 2-4, 34013 Trieste, Italy}

\date{\today}

\maketitle

\begin{abstract}
The information that a pattern of firing in the output layer of a
feedforward network of threshold-linear neurons conveys about the
network's inputs is considered. A replica-symmetric solution is found to
be stable for all but small amounts of noise. The region of instability
depends on the contribution of the threshold and the sparseness: for
distributed pattern distributions, the unstable region extends to higher
noise variances than for very sparse distributions, for which it is almost
nonexistant.
\end{abstract}


\section{Introduction}

Advances in techniques for the formal analysis of neural networks
\cite{Ami+87,Gar88,Bia+88,Tre90,Nad+93} offer insight into the behaviour of
models of biological interest. Of particular interest are methods which
allow the calculation of the information that can be conveyed by a given
neural structure, as these offer both useful intuitions and the prospect
of conducting pertinent experiments\cite{Tre+96}. The replica trick
\cite{Mez+87} has been used to achieve this in the case of binary units
\cite{Nad+93} and threshold-linear units \cite{Tre95,Sch+97a}, by
appealing to an assumption of replica symmetry. In the case of binary
units with continuous inputs, the validity of the replica-symmetric ansatz
is justified by the duality with the Gardner calculation of the storage
capacity for continuous couplings \cite{Gar88,Nad+93,Nad+94}. We now
analyse the stability of the replica symmetric solution for mutual
information in a network of threshold-linear units.

The model describes a feedforward network of threshold-linear units with
partially diluted connectivity. This is a simpler version of the
calculation described in \cite{Tre95,Sch+97a}. In the calculation
considered here, there is only one mode of operation (which we might call
``transmission''), as opposed to the division into storage and recall
modes in that calculation. There are $N$ cells in the input layer, and $M$
(proportional to $N$) in the output layer. The limit of interest is $N \to
\infty$.

$\{\eta_i\}$ are the firing rates of each cell $i$ in the input
layer. The probability density of finding a given firing
pattern is taken to be:
\begin{equation}
P(\{\eta_i\}) \prod_i d \eta_i  = \prod_i P_{\eta}(\eta_i) d\eta_i
\end{equation}
Each input cell is thus assumed to code independent information. 

$\{\xi_j\}$ are the firing rates produced in each cell in the output
layer. They are determined by the matrix multiplication of the pattern
$\{\eta_i\}$ with the synaptic weights $J_{ij}$, followed by Gaussian
distortion, thresholding and rectification.
\eq
\xi_j = \left[ \xi_0 +\sum_i c_{ij} J_{ij} \eta_i +\epsilon_j
\right]^+ = \left[{\tilde{\xi}}_j\right]^+
\label{eq:constr}
\en
\eq
\left<(\epsilon_j)^2\right>  =  \sigma^2_{\epsilon} \nonumber\\
\en
Each output cell receives $C_j$ (which we will take to be of the order of
$10^4$) connections from input layer cells:
\eq
c_{ij} \in \{0,1\} \qquad
\left< c_{ij} \right> N = C_j \qquad (C\equiv \left<C_j\right>)
\en

The mean value across all patterns of each synaptic weight is taken to be
equal across synapses, and is therefore taken into the threshold term. The
synaptic weights $J_{ij}$ are thus of zero mean, and variance $\sigma^2_J$
(all that affects the calculation is the first two moments of their
distribution).
\eq
\left<(J_{ij})^2\right> = \sigma^2_J.
\en

The average of the mutual information
\begin{equation}
I(\{\eta_i\},\{\xi_j\})=\int \prod_i d\eta_i \int \prod_j d\xi_j P(\{\eta_i\},
\{\xi_j\}) \ln {P(\{\eta_i\},\{\xi_j\})\over P(\{\eta_i\})P(\{\xi_j\})}
\end{equation}
over the quenched variables $c_{ij}, J_{ij}$ is written using the replica
trick as 
\begin{equation}
\left< I(\eta,\xi) \right>_{c,J}=\lim_{n\to 0} {1\over n} \left< \int  d\eta d\xi P(\eta,\xi)
\left\{ \left[ P(\eta,\xi) \over P(\eta) \right]^n - \left[ P(\xi)\right]^n
\right\} \right>_{c,J}.\label{eq:shaninf}
\end{equation}

The calculation is valid only for non-zero noise variance, and it will be
seen that the only region in which the solution is not well behaved is that
of very low noise variance.

\section{Calculation of mutual information}

First, introducing replica indices $\alpha = 1,..,n+1$, and breaking the
integral over $\xi$ into subthreshold and suprathreshold components, we
observe that 
\eqa
\left< I(\eta,\xi) \right>_{c,J}&=&\lim_{n \to 0} \frac{1}{n} \Bigg\{
\int d\eta \left[ \frac{1}{P(\eta)} \right]^n \prod_\alpha \left(
\int_{-\infty}^0 d{\tilde{\xi}}^\alpha \left< 
P(\eta^\alpha,{\tilde{\xi}}^\alpha)|_{\eta^\alpha=\eta} \right>_{c,J} \right)
\nonumber\\ 
&&+ \int d\eta \left[ \frac{1}{P(\eta)} \right]^n \int_0^\infty d{\tilde{\xi}}
\prod_{\alpha} \left<
P(\eta^\alpha,{\tilde{\xi}}^\alpha)|_{\eta^\alpha=\eta,{\tilde{\xi}}^\alpha={\tilde{\xi}}} \right>_{c,J}
\nonumber\\
&&- \prod_\alpha \left( \int_{-\infty}^0 d{\tilde{\xi}}^\alpha
\int  d\eta^\alpha \left<
P(\eta^\alpha,{\tilde{\xi}}^\alpha) 
\right>_{c,J}\right)  \nonumber\\
&&- \int_{0}^{\infty} d{\tilde{\xi}}
\prod_{\alpha} \left( \int d\eta^\alpha \left<
P(\eta^\alpha,{\tilde{\xi}}^\alpha)|_{{\tilde{\xi}}^\alpha={\tilde{\xi}}} 
\right>_{c,J} \right) 
\Bigg\}
\ena 
This allows us to treat both terms of Eq.~\ref{eq:shaninf} in the
same manner. To obtain the probability density $\left<
P(\eta^\alpha,{\tilde{\xi}}^\alpha)\right>$, we 
use Dirac delta functions to implement the constraints defined by
(\ref{eq:constr}):
\begin{eqnarray}
\left< P(\eta^\alpha,{\tilde{\xi}}^\alpha)\right>_{c,J} &=&
\Bigg< \int \left[ \prod_{j} 
D \left(\frac{\epsilon_j^\alpha}{\sigma_\epsilon}\right) \right]\left[
\prod_{ij} D 
\left(\frac{J_{ij}}{\sigma_J}\right) \right] \prod_{j} \delta 
\left[ {\tilde{\xi}}_j^\alpha - \xi_0 - \sum_i c_{ij} J_{ij} \eta_i^\alpha -
\epsilon_j^\alpha \right] \nonumber\\ 
& &\times P(\{\eta_i^\alpha\})^{n+1} \Bigg>_{c}
\end{eqnarray}
where
\begin{equation}
Du = \frac{du}{\sqrt{2 \pi}} e^{-u^2/2 }
\end{equation}
Using the integral form of the Dirac delta function introduces a Lagrange
multiplier $x_j^\alpha$. The integrals over the noise and interaction
distributions are performed, and the quenched average over the connections
performed in the thermodynamic limit, so that
\begin{eqnarray}
\left< P(\eta,{\tilde{\xi}})^{n+1}\right> & = & \int ( \prod_{j\alpha}
\frac{dx_j^\alpha}{2 \pi} )
\exp\left\{ i \sum_{j\alpha} x_j^\alpha ( {\tilde{\xi}}_j^\alpha - \xi_0)
\right. \nonumber \\
& & \left. - \frac{1}{2}
\sum_{j\alpha\beta} x_j^\alpha x_j^\beta \left[ \sigma_\epsilon^2
\delta_{\alpha\beta} + \frac{\sigma_J^2 C}{N} \sum_i \eta_i^\alpha
\eta_i^\beta \right] P(\{\eta_i^\alpha\})^{n+1} \right\}
\end{eqnarray}
where $\delta_{\alpha\beta}$ is the Kronecker delta. A Lagrange multiplier 
\eq
z^{\alpha\beta} = \frac{1}{N} \sum_{i} \eta_i^\alpha \eta_i^\beta 
\en
is introduced using the integral form of the Dirac delta function via an
auxiliary variable $\tilde{z}^{\alpha\beta}$. We then obtain
\eqa
 \left< P(\eta,{\tilde{\xi}})^{n+1}\right> & = & \int ( \prod_\alpha \frac{dz^\alpha
d\tilde{z}^\alpha}{2\pi/N} )( \prod_{(\alpha\beta)} \frac{dz^{\alpha\beta}
d\tilde{z}^{\alpha\beta}}{2\pi/N} ) \exp \Bigg\{ iN\sum_\alpha
z^{\alpha} {\tilde{z}}^{\alpha} + iN\sum_{(\alpha\beta)}
z^{\alpha\beta} {\tilde{z}}^{\alpha\beta} \nonumber\\
 & - & \left. \sum_{\alpha} {\tilde{z}}^{\alpha} \sum_i
(\eta_i^\alpha)^2 - \sum_{(\alpha\beta)} {\tilde{z}}^{\alpha\beta} \sum_i
\eta_i^\alpha \eta_i^\beta - \frac{1}{2} \sum_{j\alpha\beta} ({\tilde{\xi}}_j^\alpha
- \xi_0) E_{\alpha\beta} ({\tilde{\xi}}_j^\beta - \xi_0) \right.\nonumber\\
 & - & \frac{1}{2} {\rm{Tr}} \ln {\mathbf{M}} \Bigg\}
(2\pi)^{-\frac{n+1}{2}} P(\{\eta_i^\alpha\})^{n+1}
\ena
where $\mathbf{M} = \sigma_\epsilon^2 \mathbf{I} + \sigma_J^2
\mathnormal{C} \mathbf{Z}$ 
and $\mathbf{E} = {\mathbf{M}}^{-1}$. $\mathbf{Z}$ is the matrix with
elements $z^{\alpha\beta}$, and $(\alpha\beta)$ is the pair $\alpha\beta$,
$\alpha \neq \beta$.

Thus
\eqa 
 \left< I \right> = 
\lim_{n \to 0} \Bigg\{ 
\int ( \prod_\alpha\frac{dz^\alpha d{\tilde{z}}^\alpha}{2\pi/N} ) 
\exp \left[ 
iN\sum_\alpha z^\alpha {\tilde{z}}^\alpha 
- N{\mathcal{H}}_A({\tilde{z}}^\alpha) 
- M {\mathcal{G}}(z^\alpha,z^\alpha)
\right] \nonumber\\
- \int ( \prod_\alpha \frac{dz^\alpha d{\tilde{z}}^\alpha}{2\pi/N})
( \prod_{(\alpha\beta)}
\frac{dz^{\alpha\beta}d{\tilde{z}}^{\alpha\beta}}{2\pi/N} ) 
\exp \Bigg[ 
iN\sum_\alpha z^\alpha {\tilde{z}}^\alpha 
+ iN\sum_{(\alpha\beta)} z^{\alpha\beta} {\tilde{z}}^{\alpha\beta} 
\nonumber\\
- N {\mathcal{H}}_B({\tilde{z}}^\alpha,{\tilde{z}}^{\alpha\beta}) 
- M{\mathcal{G}}(z^\alpha,z^{\alpha\beta}) 
\Bigg]\Bigg\} 
\label{eq:geni}
\ena
where
\eq
{e^{-{\mathcal{H}}_A}}({\tilde{z}}^\alpha) 
= \int_\eta d\eta P(\eta) \exp \left( -\sum_\alpha {\tilde{z}}^\alpha
\eta^2 \right) \\ 
\en
\eq
{e^{-{\mathcal{H}}_B}}({\tilde{z}}^\alpha,{\tilde{z}}^{\alpha\beta})
= \int_\eta (\prod_\alpha d\eta^\alpha P(\eta^\alpha)) 
\exp{\left(
-\sum_\alpha {\tilde{z}}^\alpha(\eta^\alpha)^2
-\sum_{(\alpha\beta)}{\tilde{z}}^{\alpha\beta}\eta^\alpha\eta^\beta
\right)} \\
\en
\eqa
{e^{-\mathcal{G}}}(z^\alpha,z^{\alpha\beta})
&=& e^{-\frac{1}{2}\mathrm{Tr}\ln{\mathbf{M}}}
\Bigg\{
\int_0^\infty \frac{d{\tilde{\xi}}}{\sqrt{2\pi}} 
\exp{ -\frac{1}{2} ({\tilde{\xi}}-\xi_0)^2 \sum_{\alpha\beta} E_{\alpha\beta} }
\nonumber\\
&+& \int_{-\infty}^0 (\prod_\alpha \frac{d{\tilde{\xi}}^\alpha}{\sqrt{2\pi}})
\exp\left[
-\frac{1}{2}\sum_{\alpha\beta} ({\tilde{\xi}}^\alpha-\xi_0) E_{\alpha\beta}
({\tilde{\xi}}^\beta-\xi_0) \right]
\Bigg\}
\ena

\section{Replica symmetric solution}

The assumption of replica symmetry can be written
\eqa
z_A^\alpha = z_A^{\alpha\beta} &=& z_{0A}(n) \qquad 
i {\tilde{z}}_A^{\alpha} = {\tilde{z}}_{0A}(n) \nonumber\\
z_B^\alpha &=& z_{0B}(n) \qquad
i {\tilde{z}}_B^{\alpha} = {\tilde{z}}_{0B}(n)\nonumber\\
z_B^{\alpha\beta} &=& z_{1}(n) \qquad
i {\tilde{z}}_B^{\alpha\beta} = - {\tilde{z}}_{1}(n) \nonumber\\
\ena
The saddle-point method is utilized in the thermodynamic limit, yielding
the saddle-point equations
\ml
\eq
z_{0A} = \left< \eta^2 \right>_\eta
\en
\eq
{\tilde{z}}_{0A} = 0
\en
\eq
z_{0B} = \left< \eta^2 \right>_\eta
\en
\eq
{\tilde{z}}_{0B} = 0
\en
\eq
z_{1} = - \int_{-\infty}^{\infty} Ds
\left< (\eta^2+\frac{s\eta}{\sqrt{{\tilde{z}}_{1}}})
\exp(-\frac{{\tilde{z}}_{1}}{2} \eta^2 - s \sqrt{{\tilde{z}}_{1}}
\eta)\right>_\eta 
\ln \left< \exp(-\frac{{\tilde{z}}_{1}}{2}\eta^2 - s \sqrt{{\tilde{z}}_{1}}
\eta)\right>_\eta
\label{eq:z1b}
\en
\eqa
{\tilde{z}}_{1} &=&- \sigma_J^2Cr \Bigg\{ \frac{\xi_0}{(p_B+q_B)^{3/2}}
\sigma(\frac{\xi_0}{\sqrt{p_B+q_B}})
- \frac{1}{p_B} \phi\left(\frac{\xi_0}{\sqrt{p_B+q_B}}\right) \nonumber\\
&+& \int_{-\infty}^\infty Dt \left[ 1 + \ln \phi\left(
\frac{-\xi_0-t\sqrt{q_B}}{\sqrt{p_B}} \right) \right] \sigma \left(
\frac{-\xi_0-t\sqrt{q_B}}{\sqrt{p_B}} \right) p_B^{-3/2}
\left(\xi_0+\frac{t(p_B+q_B)}{\sqrt{q_B}} \right)
\Bigg\}
\label{eq:z1bt}
\ena
\eml
and the expression for the information per input cell
\eqa
 \left< i \right> =&r& G(p_A,q_A) + \frac{1}{2} z_{1}
{\tilde{z}}_{1} - r G(p_B,q_B)\nonumber\\
&-& \int_{-\infty}^{\infty} Ds
\left< \exp(-\frac{1}{2} {\tilde{z}}_1 \eta^2 - s \sqrt{{\tilde{z}}_1}
\eta)\right>_\eta 
\ln \left< \exp(-\frac{1}{2} {\tilde{z}}_1 \eta^2 - s \sqrt{{\tilde{z}}_1}
\eta)\right>_\eta
\ena
where
\eqa
 G(p,q) = \frac{p\xi_0}{2(p+q)^{3/2}} \sigma \left(
\frac{\xi_0}{\sqrt{p+q}} \right) - \frac{1}{2}(1+\ln{p}) {\phi\left(
\frac{\xi_0}{\sqrt{p+q}}\right)} \nonumber\\
+ \int_{-\infty}^{\infty} Dt \phi\left( 
\frac{-\xi_0-t\sqrt{q}}{\sqrt{p}} \right) 
\ln \phi\left( \frac{-\xi_0-t\sqrt{q}}{\sqrt{p}} \right) 
\label{eq:Grs}
\ena
and
\eqa
\left< x(\eta) \right>_\eta &=& \int_\eta d\eta P(\eta) x(\eta) \nonumber\\
\phi(x) &=& \int_{-\infty}^x Ds \nonumber\\
\sigma(x) &=& \frac{1}{\sqrt{2 \pi}} e^{-x^2/2} \nonumber\\
\nonumber
\ena
\eqa
p_A &=& \sigma_\epsilon^2 \qquad\qquad p_B = \sigma_\epsilon^2 + \sigma_J^2 C
(z_{0B}-z_{1}) \nonumber\\
q_A &=& \sigma_J^2 C z_{0A} \qquad q_B = \sigma_J^2 C z_{1} .\nonumber\\
\label{eq:pq}
\ena
We refer to $r = M/N$ as the anatomical divergence. 

This expression must in general be evaluated numerically. However,
considering some limiting cases can give us some insight into the
behaviour of the solution. In particular, the limit of linear processing
can be obtained by taking $\xi_0 \to +\infty$. In this limit,
Eq.~\ref{eq:z1bt} reduces to
\eq
{\tilde{z}}_{1} \to \frac{\sigma_J^2Cr}{p_B} .
\en
The information per neuron obtained in the linear limit is
\eqa
 \left<i\right> \to &&\frac{1}{2} r \ln \frac{p_B}{p_A} + \frac{1}{2}
z_{1}{\tilde z}_{1} \nonumber\\
&-& \int_{-\infty}^{\infty} Ds
\left< \exp(-\frac{1}{2} {\tilde{z}}_1 \eta^2 - s \sqrt{{\tilde{z}}_1}
\eta)\right>_\eta 
\ln \left< \exp(-\frac{1}{2} {\tilde{z}}_1 \eta^2 - s \sqrt{{\tilde{z}}_1}
\eta)\right>_\eta .
\ena

The information obtained in this limit is bounded by that which would be
obtained from a simple Gaussian channel calculation, where we consider the
channel
\eq
\xi_j^* = \sum_i c_{ij}J_{ij}\eta_i + \epsilon_j,
\en
and perform the annealed and quenched averages to obtain the signal
variance $\sigma_J^2C(\left<\eta^2 \right>_\eta - 
\left<\eta\right>_\eta^2)$, and information per input cell
\eq
I_{gauss} = \frac{r}{2} \ln \left[1+
\frac{\sigma_J^2C(\left<\eta^2\right>_\eta -
\left<\eta\right>_\eta^2)}{\sigma_\epsilon^2} \right] .
\en
The Gaussian channel information provides an upper limit corresponding to
the optimal $\eta$ distribution (for transmitting maximal information
given a constraint on the signal power), and no dependence upon the same
inputs of the output cells.

Within the linear limit, we can consider the special case of high 
noise variance (low signal to noise ratio). As $\sigma_\epsilon^2 \to
\infty$,  
\eq
{\tilde{z}}_{1} \sim \frac{\sigma_J^2Cr}{\sigma_\epsilon^2},
\en
and
\eq
z_{1} \simeq \left< \eta \right>^2 + O({\tilde{z}}_{1}) .
\en 
The information therefore falls to zero as
\eq
\left< i \right> \sim \frac{\sigma_J^2 C r (\left<
\eta^2 \right>_\eta - \left<\eta\right>_\eta^2)}{2 \sigma_\epsilon^2} ,
\en
i.e. inversely with noise variance, as one would expect. We thus can see
that for linear neurons with low signal to noise ratio, the transmitted
information approaches the Gaussian channel limit. \footnote{It can also
be shown (we have done so for the case of a Gaussian $\eta$ distribution),
that as $r \to 0$, the Gaussian channel bound is also reached.}

\begin{figure}
\begin{center}
\epsfxsize=10cm
\epsfysize=8.6cm
\leavevmode
{\bfseries\Large a\epsffile{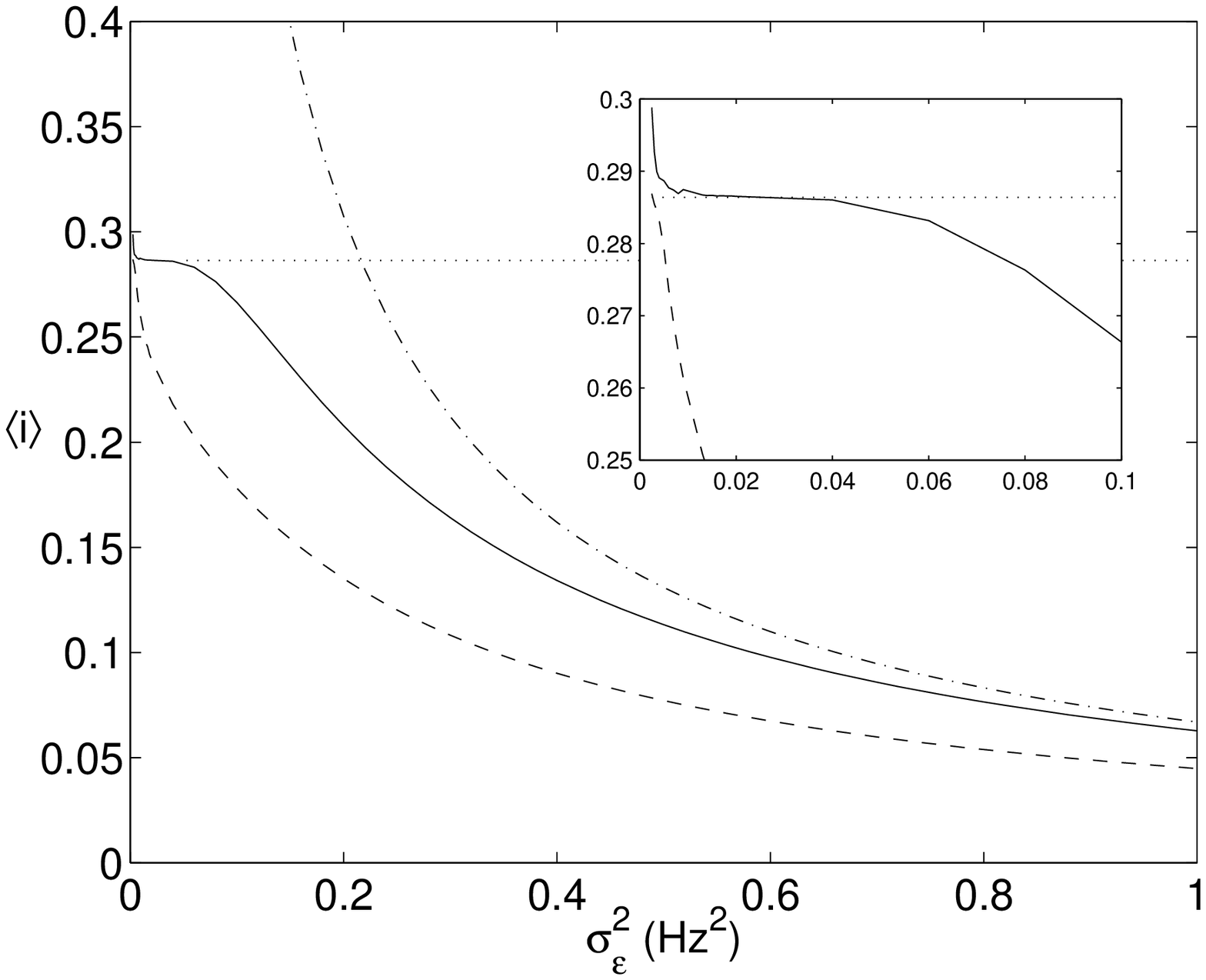}}
\end{center}
\begin{center}
\epsfxsize=10cm
\epsfysize=8.6cm
\leavevmode
{\bfseries\Large b\epsffile{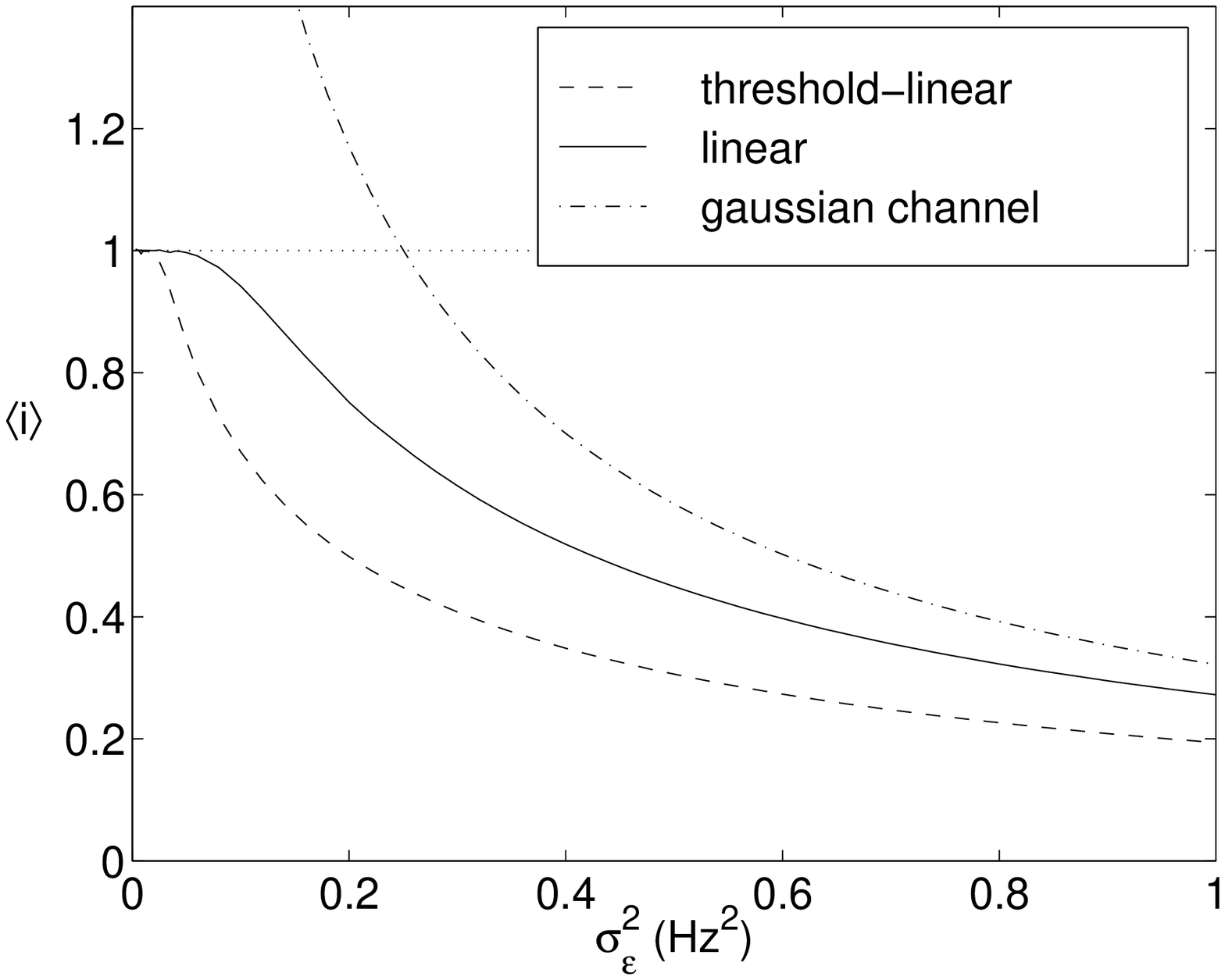}}
\end{center}
\caption{Mutual information, measured in bits, as a function of
noise variance. The dashed line is for a threshold $\xi_0=-0.4$, whereas
the solid line is for the limit of linear neurons. The dot-dashed line
indicates the simple gaussian channel for comparison. The entropy of the
input pattern distribution is indicated by the horizontal dotted line.
(a) Input pattern distribution sparseness of 0.05. (b)
Sparseness of 0.50.}
\label{fig:infcf}
\end{figure}

The numerical solution of the mutual information expression, as a function
of the noise variance, is shown in Fig.~\ref{fig:infcf}, both for the
case of linear units and for units with a threshold of $\xi_0 = -0.4$,
representing threshold-linear behaviour. This is shown for a
binary pattern distribution of sparseness $a$, where the sparseness of a
distribution is a mean-invariant measure of spread and is defined in general as
\eq
a = \frac
{\left< \eta \right>_\eta^2}
{\left< \eta^2 \right>_\eta} .
\en
This measure is `more sparse' for smaller $a$, and reduces to the fraction
of units `on' in the case of a binary distribution. The Gaussian channel
bound appears on the same graphs for comparison.

The mutual information should be bounded by the pattern entropy as the
noise variance becomes very small. As the noise variance decreases, the
replica-symmetric solution approaches this bound in both the linear and
threshold-linear cases. It can be seen, however, that for very small noise
variances, the replica-symmetric solution changes direction and crosses
this physical boundary. Inspection of Eq. \ref{eq:Grs} reveals
divergence of the mutual information solution in the limit
$\sigma_\epsilon^2 \to 0$; this is in keeping with our intuition from the
beginning that the calculation should not be valid in the deterministic
limit. However, for such low noise variance the information has
essentially saturated in any case. For threshold-linear neurons, the
solution is also unstable to replica-symmetry-breaking fluctuations for
relatively low noise variance, as will be discussed in the next section.

\section{Stability of the replica-symmetric solution}

The stability of the replica-symmetric solution is analysed after the
style of de Almeida-Thouless \cite{Alm+78}. For the solution for free
energy this was addressed in the context of Hopfield-Little type
autoassociative neural networks in \cite{Ami+87}, and for an
autoassociator with threshold-linear units and for a threshold-linear
variant of the Sherrington-Kirkpatrick model in \cite{Tre91}. For the
solution for another quantity, the Gardner volume, this was addressed in
\cite{Gar88} for Ising ($\pm 1$) neurons. In contrast, here we are
determining the stability of the solution for mutual information in a
network comprised of threshold-linear neurons, although the technique
proceeds very similarly.

Fluctuations in the transverse (replica-symmetry breaking, RSB) and
longitudinal (replica-symmetric, RS) directions are decoupled, and hence
can be analysed separately. Longitudinal fluctuations can be disregarded
\cite{Alm+78,Gar+88} if a unique saddle-point is obtained, which appears
to be the case. We will therefore concentrate upon transverse
fluctuations.

We wish to consider small deviations in the saddle-point parameters about
the replica-symmetric saddle-point,
\eqa
z^{\alpha\beta} = z_{1} + \delta z^{\alpha\beta} \nonumber\\
{\tilde{z}}^{\alpha\beta} = {\tilde{z}}_{1} + \delta
{\tilde{z}}^{\alpha\beta}
\ena
Quadratic fluctuations in the function
\eq
{\mathcal{B}}(z^\alpha, {\tilde{z}}^\alpha, z^{\alpha\beta},
{\tilde{z}}^{\alpha\beta}) = iN\sum_\alpha z^\alpha {\tilde{z}}^\alpha  
+ iN\sum_{(\alpha\beta)} z^{\alpha\beta} {\tilde{z}}^{\alpha\beta} 
\nonumber\\
- N {\mathcal{H}}_B({\tilde{z}}^\alpha,{\tilde{z}}^{\alpha\beta}) 
- M{\mathcal{G}}(z^\alpha,z^{\alpha\beta}) .
\en
give us the stability matrix 
\eq
\mathbf{\Gamma} = \left[ \matrix{ 
\frac{\displaystyle\partial^2 \mathcal{B}}{\displaystyle\partial
z^{\alpha\beta} \partial z^{\gamma\delta}} &
\frac{\displaystyle\partial^2 \mathcal{B}}{\displaystyle\partial
z^{\alpha\beta} \partial (i{\tilde{z}}^{\gamma\delta})}\cr
\frac{\displaystyle\partial^2 \mathcal{B}}{\displaystyle\partial
(i{\tilde{z}}^{\alpha\beta}) \partial z^{\gamma\delta}} &
\frac{\displaystyle\partial^2 \mathcal{B}}{\displaystyle\partial
(i{\tilde{z}}^{\alpha\beta}) \partial (i{\tilde{z}}^{\gamma\delta})}
} \right] =
\left[ \matrix{
A^{(\alpha\beta)(\gamma\delta)} & \delta_{(\alpha\beta)(\gamma\delta)} \cr
\delta_{(\alpha\beta)(\gamma\delta)} & B^{(\alpha\beta)(\gamma\delta)} }
\right]
\label{eq:stabmat}
\en
where $\delta_{(\alpha\beta),(\gamma\delta)} =
\delta_{\alpha\gamma}\delta_{\beta\delta} +
\delta_{\alpha\delta}\delta_{\beta\gamma}$.
In constrast to previous calculations based on quantities such as
free energy, the expression for mutual information involves $n+1$
replicas. There are $n(n+1)/2$ independent variables $z^{\alpha\beta}$,
and the same number of independent ${\tilde{z}}^{\alpha\beta}$.
$\mathbf{\Gamma}$ is thus an $n(n+1) \times n(n+1)$ matrix. 

The transverse eigenvalues of this matrix are given by the eigenvalues
of the matrix
\eq
\left( \matrix{
\lambda_A & 1 \cr
1 & \lambda_B
} \right) ,
\en
where $\lambda_A$ and $\lambda_B$ are the transverse eigenvalues of
the submatrices $A^{(\alpha\beta)(\gamma\delta)}$ and
$B^{(\alpha\beta)(\gamma\delta)}$ respectively. Calculation of these
involves consideration of the symmetry properties of the submatrices, and
is detailed in the Appendix. The eigenvalue equations reduce to
\eqa
&\lambda_A&+c = \lambda \nonumber\\
&1&+c\lambda_B = c\lambda 
\ena
We thus have the two replicon mode eigenvalues
\eq
\lambda_\pm = \frac{1}{2} (\lambda_A+\lambda_B) \pm \sqrt{ \frac{1}{4}
(\lambda_A - \lambda_B)^2 + 1}
\label{eq:lampm}
\en

For stability, the product of the eigenvalues must be non-negative. A
further subtlety is introduced here. $\lambda_+$ can be seen to be $> 0$
irrespective of $\sigma_\epsilon^2$ or $a$. $\lambda_-$, on the other
hand, changes sign, moving from negative to positive for smaller
$\sigma_\epsilon^2$. However, intuitively we expect, from the analogy of
the noise with the `temperature' parameter in other models of neural
networks\cite{Ami+87} and physical systems\cite{She+75} that if
replica-symmetry breaking is to set in, it will do so at low noise
variances. This is confirmed by the eminently sensible behaviour of the
mutual information curves of Fig.~\ref{fig:infcf} at medium to high
noise, but nonphysical behaviour at very low noise values. It can be
concluded that, as occurs in \cite{Ami+87,Tre91}, a sign reversal has been
introduced due to the integration contour, which must be corrected. 

\begin{figure}
\begin{center}
\epsfxsize=6.8cm
\epsfysize=6cm
\leavevmode
\parbox[t]{7.5cm}{\bfseries\Large a\epsffile{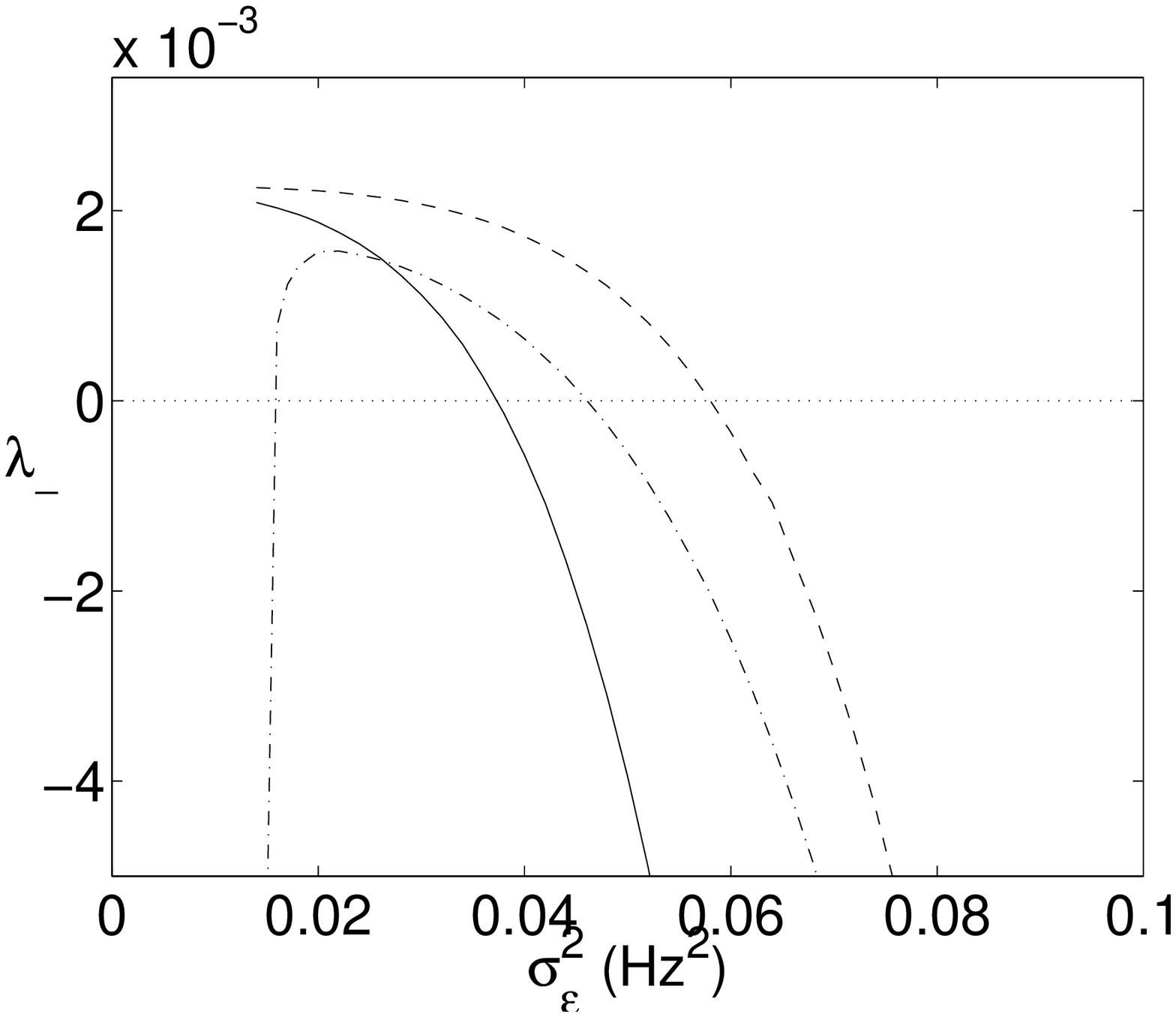}}
\epsfxsize=6.8cm
\epsfysize=6cm
\leavevmode
\parbox[t]{7.5cm}{\bfseries\Large b\epsffile{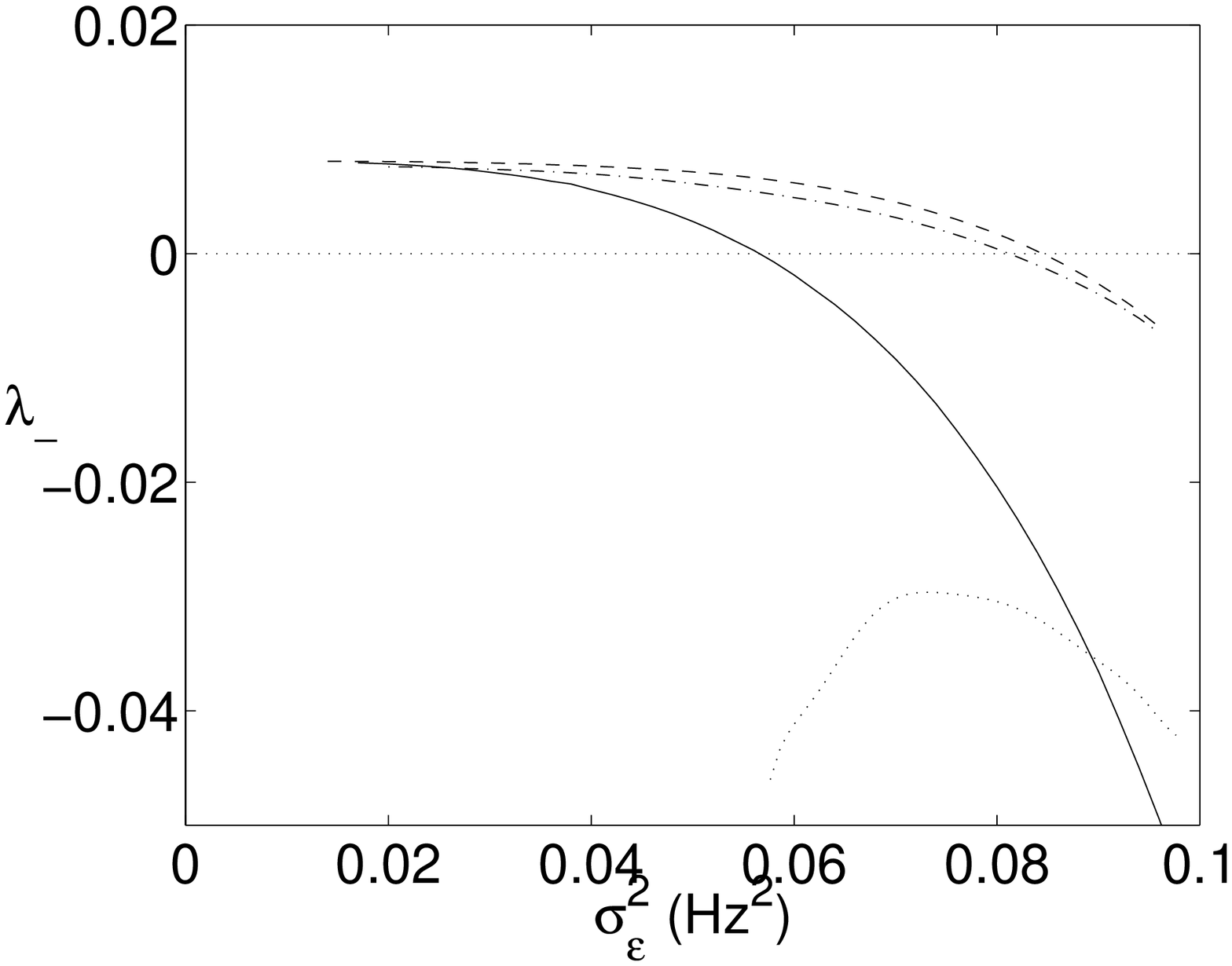}}
\epsfxsize=6.8cm
\epsfysize=6cm
\leavevmode
\parbox[t]{7.5cm}{\bfseries\Large c\epsffile{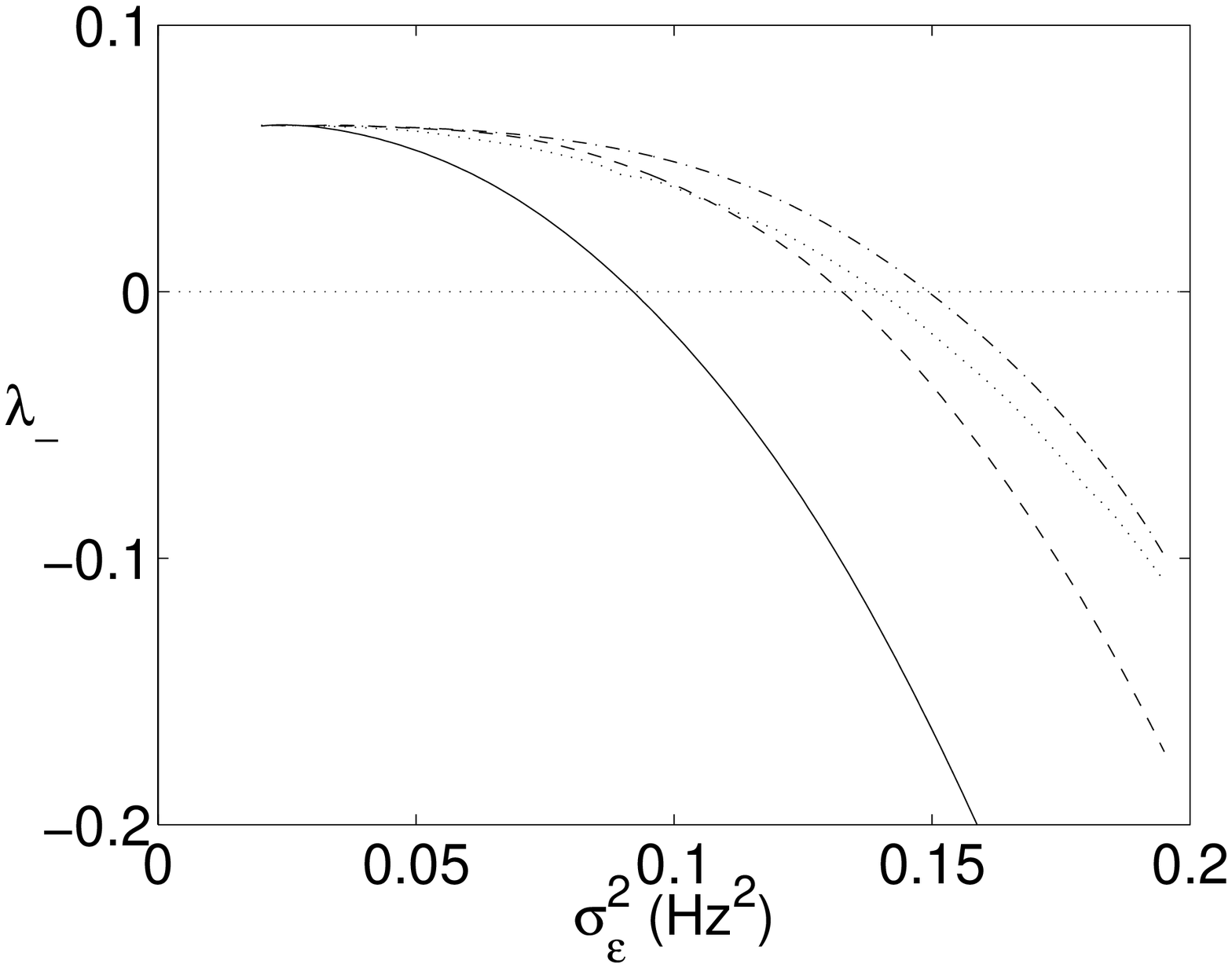}}
\end{center}
\caption{The behavior of the replicon mode eigenvalue $\lambda_-$ as a
function of noise variance. (a) Input sparseness $a=0.05$ (b) $a=0.10$ (c)
$a=0.50$. In each of these graphs the solid line indicates the eigenvalue
for threshold $\xi_0 = -0.4$, the dashed curve $\xi_0 = 0.0$, the
dot-dashed curve $\xi_0=0.4$, and the dotted curve $\xi_0=0.8$. The
replica symmetric solution is unstable in regions where these curves lie
above the horizontal dotted line. In case (a), the $\xi_0=0.8$ line lies
below the region examined in the graph. }
\label{fig:eig}
\end{figure}

These equations have been numerically solved for
$\lambda_-$. Fig.~\ref{fig:eig} shows the behavior of $\lambda_-$ for a
range of sparsenesses and thresholds. Where the eigenvalue passes above
the zero axis (dotted line), a phase of RS-instability is
indicated. Fig.~\ref{fig:eig}a is for the situation of quite sparse coding
of the patterns. As the noise is reduced from the high noise region, in
which the RS solution is stable, the eigenvalue changes sign, and an
unstable region is entered. In the case of threshold $\xi_0$ = 0.4, which
represents only a very small degree of threshold-like behavior, the
eigenvalue can be seen to curve back and change sign again at lower noise
values still. Due to non-convergence of numerical integration, it is not
possible to examine extremely small noise values; therefore it is not
clear from this diagram whether the eigenvalue also falls below zero
again for the other curves plotted in this figure, or if it instead has a
finite value at zero noise. However, any region of RS stability at noise
variances this low would obviously be irrelevant for the same numerical
reasons.

\begin{figure}
\begin{center}
\epsfxsize=14cm
\epsfysize=12cm
\leavevmode
\epsffile{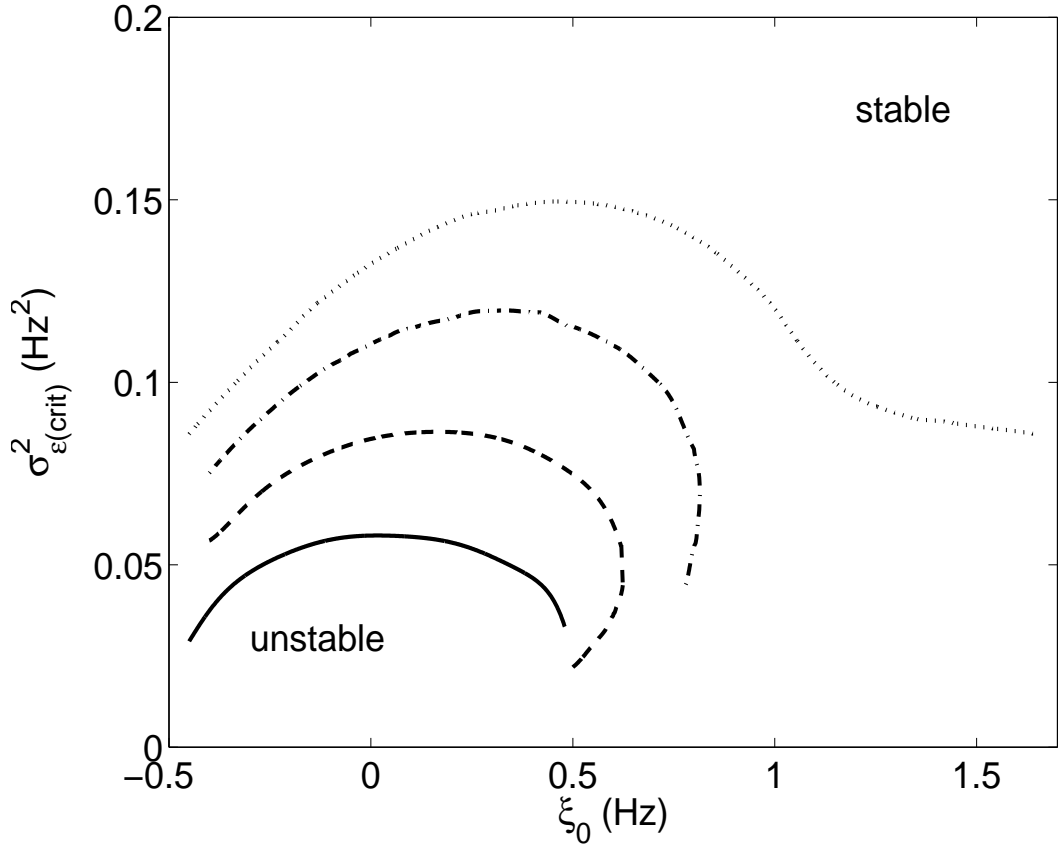}
\end{center}
\caption{A phase diagram showing the critical noise variance as a function
of the threshold parameter, $\xi_0$ -- the larger $\xi_0$ is, the more
linear the regime. Solid curve, sparseness $a=0.05$; dashed curve,
$a=0.10$; dot-dashed curve, $a=0.20$; dotted curve, $a=0.50$.}
\label{fig:thresh}
\end{figure}

It is apparent from Figs.~\ref{fig:eig}(b) and (c) that as the input
distribution is made less sparse ($a$ is increased), the critical amount of
noise below which instability arises increases. This will be discussed
again shortly. Another effect that can be seen in Figs.~\ref{fig:eig}(a)
and (b) is that, as the neurons are made more linear ($\xi_0$ is
increased), the critical noise first rises, then falls. This becomes more
clear after plotting a phase diagram of noise against $\xi_0$
(Fig.~\ref{fig:thresh}). For low $a$ (sparse distributions), the critical
noise rises, falls, and then curves back around on itself -- after the
neurons become sufficiently linear, there is no more region of
instability. As the pattern code becomes less sparse, at first the region
of instability merely expands. When $a$ reaches a certain value, however,
the edge of the unstable region no longer curls in on itself, but extends
outwards. At a sparseness of 0.5, for instance, the critical noise thus
first rises with increasing linearity, taking longer to reach its peak
than for more sparse distributions, then falls, and finally levels off and
decreases slowly. The sparseness at which this change in behavior is
exhibited is independent of the parameters of the system, and can be seen
from Fig.~\ref{fig:thresh} to lie somewhere between 0.2 and 0.5.

In the special case of the linear limit, in which $\xi_0 \to \infty$,
$\lambda_A$ disappears (see Appendix), and stability is assured. For
finite $\xi_0$ and above the coefficient of sparseness referred to in the
previous paragraph, though, there is a distinct and reasonably large
region of instability.

\begin{figure}
\begin{center}
\epsfxsize=6.8cm\epsfysize=6cm
\leavevmode
\parbox[t]{7.5cm}{\bfseries\Large a\epsffile{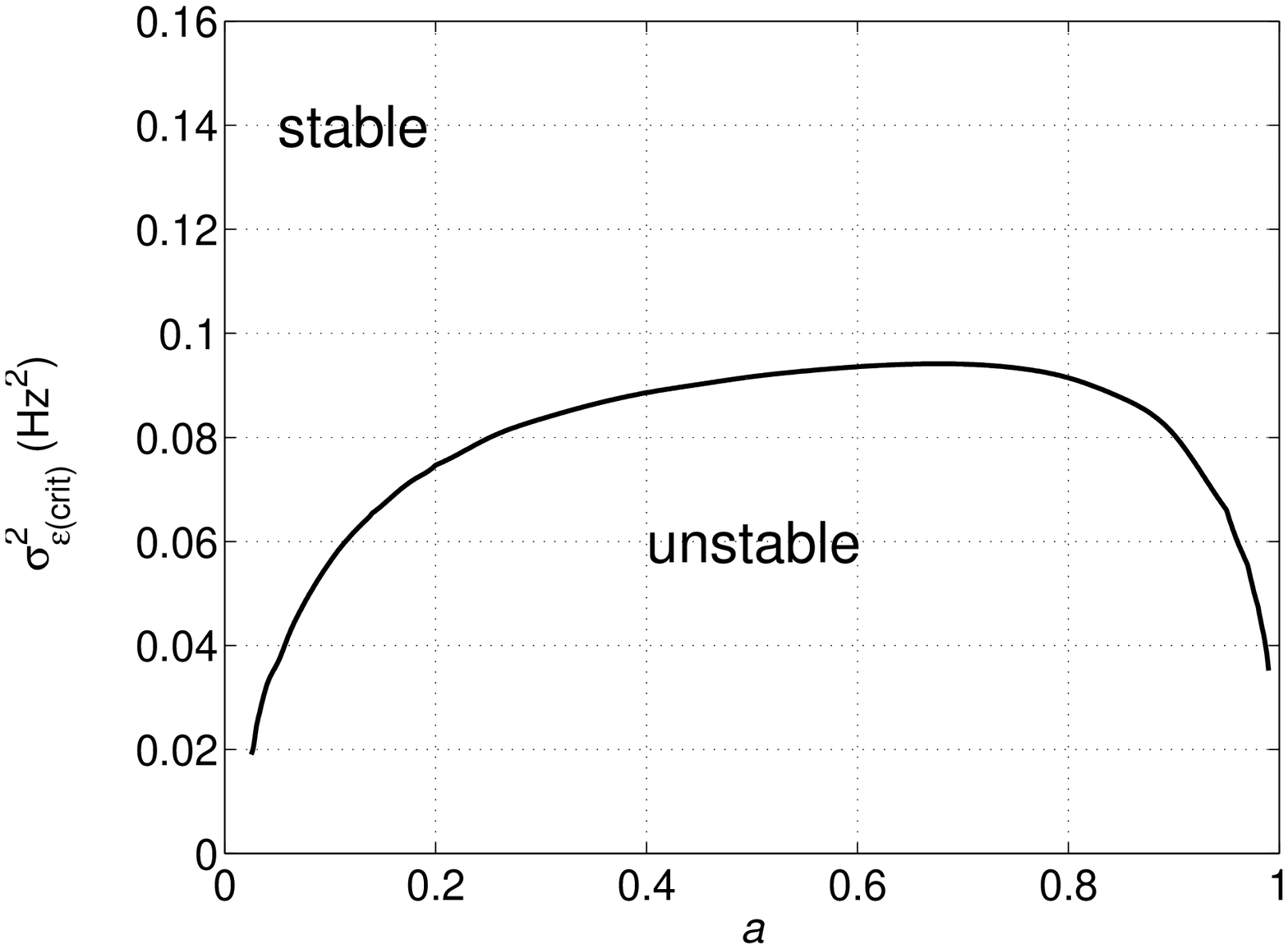}}
\epsfxsize=6.8cm
\epsfysize=6cm
\leavevmode
\parbox[t]{7.5cm}{\bfseries\Large b\epsffile{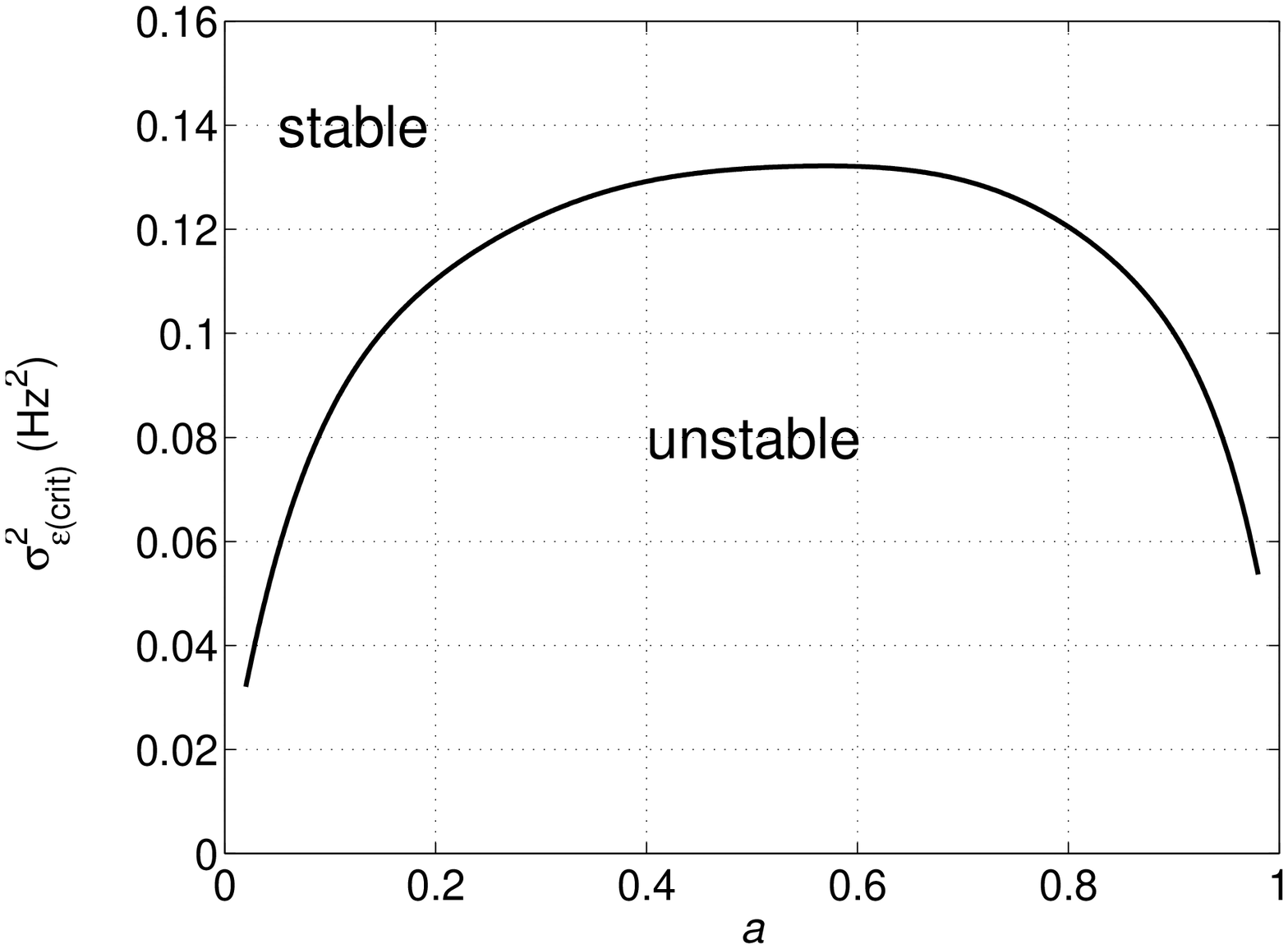}}
\end{center}
\begin{center}
\epsfxsize=6.8cm
\epsfysize=6cm
\leavevmode
\parbox[t]{7.5cm}{\bfseries\Large c\epsffile{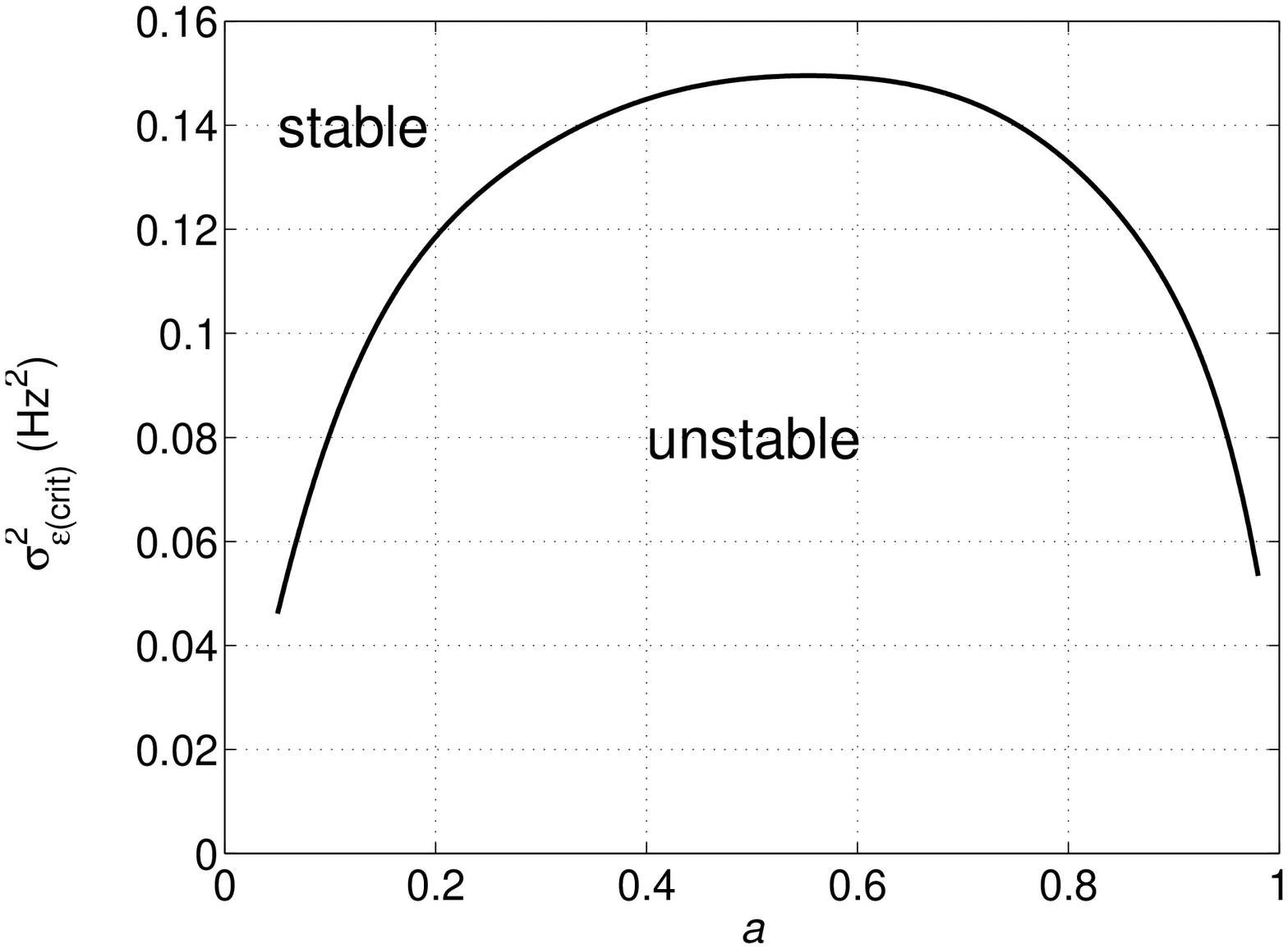}}
\epsfxsize=6.8cm
\epsfysize=6cm
\leavevmode
\parbox[t]{7.5cm}{\bfseries\Large d\epsffile{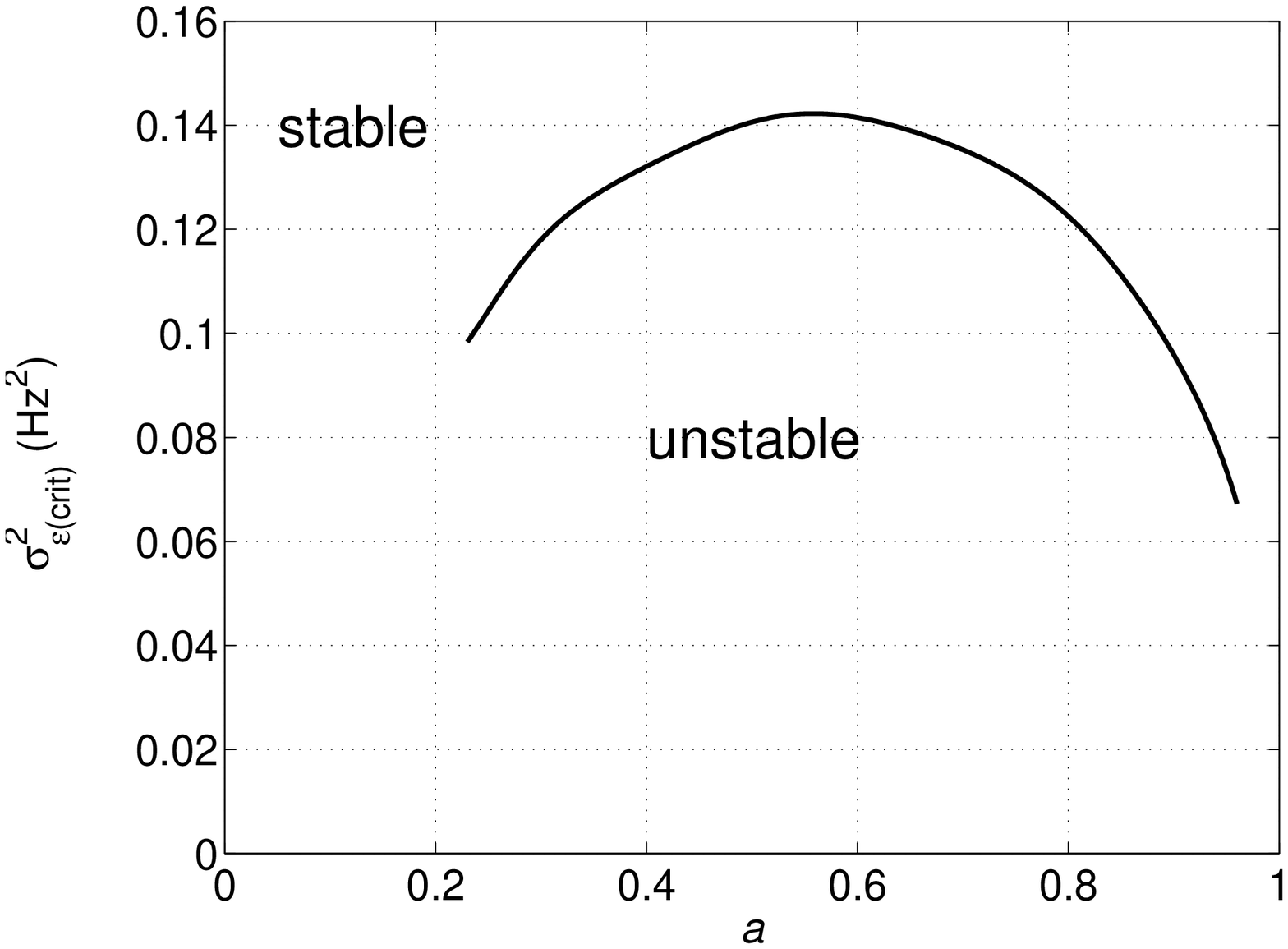}}
\end{center}
\caption{The phase diagram for information transmission, for
$r=2$ and $\sigma_J^2 = 1/C$. (a) Threshold $\xi_0=-0.4$. (b) Threshold
$\xi_0=+0.0$. (c) Threshold $\xi_0=+0.4$. (d) Threshold $\xi_0=+0.8$.}
\label{fig:phase}
\end{figure}

The resulting phase diagrams are shown in
Fig.~\ref{fig:phase}. Fig.~\ref{fig:phase}(a) shows the situation for
$\xi_0=-0.4$, which corresponds to threshold-linear behavior. As $\xi_0$ is
increased (Fig.~\ref{fig:phase}b-d; the neurons are made progressively
``more linear''), the critical noise variance at which instability of the
RS solution sets in first increases, and then decreases, as would be
expected from Fig.~\ref{fig:thresh}. In Fig.~\ref{fig:phase}(d), the line
of critical noise variance abruptly stops at $a \sim 0.23$: at
this point, the replicon-mode eigenvalue passes below the zero axis, and
stability is assured. In all cases, it is apparent that in particular for
very sparse distributions, the replica-symmetric equations are valid down
to quite low noise. For less sparse coding, where the pattern entropy is
significantly higher, the replica-symmetry-broken solution would seem to
be relevant for higher noise variances. 

\begin{figure}
\begin{center}
\epsfxsize=10cm
\epsfysize=9cm
\leavevmode
\parbox[t]{10.5cm}{\epsffile{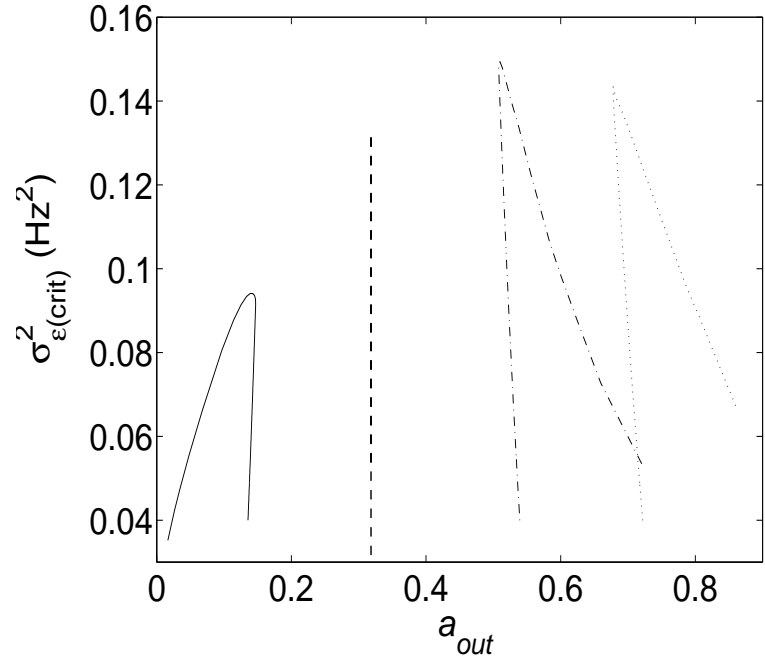}}
\end{center}
\caption{The marginal noise variance as a function of the
sparseness of the {\em output} distribution. The solid line represents the
curve for $\xi_0=-0.40$ (the same situation as Fig.~\ref{fig:phase}a), the
dashed curve $\xi_0=0.0$, the dot-dashed curve $\xi_0=+0.40$, and the
dotted curve $\xi_0=+0.80$. Note that for $\xi_0=0.0$ the output
sparseness is fixed at $1/\pi$, as explained in the text, so this
particular line is not informative about the relative region of instability.}
\label{fig:out}
\end{figure}

It should be noted that the sparseness of the distribution of outputs is
not the same as that of the inputs. This can be determined by
\eq
a_{out} = \frac{ \left<\xi\right>_{\xi^+}^2 }{ \left<\xi^2\right>_{\xi^+}}
\label{eq:aout}
\en
where
\eqa
\left<x(\xi)\right>_{\xi^+} = 
\int_0^\infty \frac{d\xi}{\sqrt{2\pi\sigma_\xi^2}} x(\xi) \exp
-\frac{(\xi-\xi_0)^2}{2\sigma_\xi^2} \nonumber\\
\sigma_\xi^2 = \sigma_\epsilon^2 + \sigma_J^2 C ( \left< \eta^2
\right>_\eta - \left< \eta \right>_\eta^2 ).
\label{eq:aout2}
\ena
The lines of marginal stability for $\xi_0=-0.4$, $\xi_0=0.0$, $\xi_0=0.4$
and $\xi_0=0.80$ are replotted in Fig.~\ref{fig:out} against the output
sparseness. Although the phase diagrams look fairly similar when
plotted as a function of input sparseness, they occupy different regions
of the output sparseness domain because of the thresholding. It
is also worth noting that because of the mapping performed by
Eq.~\ref{eq:aout}, the boundaries of the regions in Fig.~\ref{fig:phase}
do not necessarily form the boundaries of the regions in the
output-sparseness plane, which in some instances constitute points from
inside the above curves.

For neurons operating in the threshold-linear regime (left curve, $\xi_0 <
0.0$), where output sparseness is effectively constrained by the
thresholding, the stability characteristics are qualitatively as has been
described earlier. For $\xi_0 = 0.0$, it is apparent from Eqs.
\ref{eq:aout} and \ref{eq:aout2} that the output sparseness is constant
(regardless of the input sparseness) at a value of $1/\pi$. As $\xi_0$ is
increased above zero, the output becomes less sparse, and the line of
marginal stability is flipped horizontally (because in this range the
entropy is higher for smaller $a_{out}$; right curves). Assuming that the
sparseness of coding in connected sets of neurons in the brain tends to be
similar, the former curve (for threshold-linear behaviour) might be
considered the more biologically applicable, with the threshold in this
model incorporating functionally the constraint on the degree of neural
activity.

\section{Conclusions}

This paper has detailed the replica symmetric solution for the information
transmitted by a feedforward network of threshold-linear neurons, and
examined its stability to fluctuations in the direction of replica
symmetry breaking. It appears that for sparse pattern distributions,
replica-symmetry breaking only sets in at noise variances sufficiently
small that we might reasonably consider them to be `beyond the realm of
biological interest', at least for noisy cortical cells.  We believe that,
quite importantly, there is every reason to expect that these results
carry over to the slightly more complicated `Schaffer collateral'
calculation described in \cite{Tre95,Sch+97a}. There is thus reason to
feel confidence in the replica-symmetric assumption when analysing neural
networks in areas such as the hippocampus which are known to code
sparsely.

When more distributed (less sparse) encoding is used, the mutual
information solution is prone to instability to replica-symmetry-breaking
fluctuations at higher amounts of noise than in the sparse case. It is
not clear from the current analysis what the quantitative effect of broken
replica symmetry might be, or what the form of the exact solution would be
in that case (e.g. the Parisi ansatz\cite{Par80}). Care should therefore
be taken when analysing the information conveyed by networks using more
distributed encoding.

\section*{Acknowledgements}

We would like to thank S Panzeri, F Battaglia and C Fulvi-Mari for useful
discussions. In particular, we would like to thank ET Rolls for his role
in the collaborative research environment that allowed this work
to be undertaken. SS would also like to thank the Oxford McDonnell-Pew
Centre for Cognitive Neuroscience for a Research Studentship.

\newpage

\section*{Appendix}

In this appendix the transverse eigenvalues of the submatrices
$A^{(\alpha\beta)(\gamma\delta)}$ and $B^{(\alpha\beta)(\gamma\delta)}$
are calculated.  Both $A^{(\alpha\beta),(\gamma\delta)}$ and
$B^{(\alpha\beta),(\gamma\delta)}$ have three different types of matrix
elements depending on whether none, one or two replica indices of the pair
$(\alpha\beta)$ equal those of the pair $(\gamma\delta)$. The three
possible values $A^{(\alpha\beta),(\gamma\delta)}$ can take are:
\eqa
P = \frac{\partial^2 \mathcal{B}}{\partial z^{\alpha\beta} \partial
z^{\alpha\beta}} &=& \frac{\sigma_J^4 C r}{4 W} \frac{(q^2+2pq)^2}{p^4(p+q)^4}
\Bigg\{\int_0^\infty \frac{d\xi}{\sqrt{2\pi}}
(\xi-\xi_0)^4 \exp \left[ -\frac{(\xi-\xi_0)^2}{2(p+q)} \right] \nonumber\\
&+& \int_{-\infty}^\infty \frac{dt}{\sqrt{2\pi}} \bigg[ (\xi-\xi_0)^2
\bigg]^2_{\xi^-}(\textstyle\frac{1}{2},t) \Bigg\}\nonumber\\
Q = \frac{\partial^2 \mathcal{B}}{\partial z^{\alpha\beta} \partial
z^{\alpha\gamma}} &=& \frac{\sigma_J^4 C r}{4 W} \frac{(q^2+2pq)^2}{p^4(p+q)^4}
\Bigg\{\int_0^\infty \frac{d\xi}{\sqrt{2\pi}} (\xi-\xi_0)^4 \exp \left[
-\frac{(\xi-\xi_0)^2}{2(p+q)} \right]\nonumber\\ 
&+& \int_{-\infty}^\infty \frac{dt}{\sqrt{2\pi}} \bigg[ (\xi-\xi_0)^2
\bigg]_{\xi^-}(\textstyle\frac{1}{3},t) \bigg[ (\xi-\xi_0)
\bigg]^2_{\xi^-}(\textstyle\frac{1}{3},t) \Bigg\} \qquad (\beta \neq \gamma)
\nonumber\\ 
R = \frac{\partial^2 \mathcal{B}}{\partial z^{\alpha\beta} \partial
z^{\gamma\delta}} &=& \frac{\sigma_J^4 C r}{4 W} \frac{(q^2+2pq)^2}{p^4(p+q)^4}
\Bigg\{\int_0^\infty \frac{d\xi}{\sqrt{2\pi}}
(\xi-\xi_0)^4 \exp \left[ -\frac{(\xi-\xi_0)^2}{2(p+q)} \right] \nonumber\\
&+& \int_{-\infty}^\infty \frac{dt}{\sqrt{2\pi}} \bigg[ (\xi-\xi_0)
\bigg]^4_{\xi^-}(\textstyle\frac{1}{4},t) \Bigg\}
\qquad (\alpha \neq \gamma, \beta \neq \delta) ,\nonumber\\
\label{eq:Apos}
\ena
where $\bigg[ x(\xi) \bigg]_{\xi^-}(k,t)$ is defined as
\eq
\bigg[ x(\xi) \bigg]_{\xi^-}(k,t) = \int_{-\infty}^0
\frac{d\xi}{\sqrt{2\pi}} 
x(\xi) \exp \left[ -\frac{k}{2p} (\xi-\xi_0)^2 + kt\sqrt{\frac{q}{p(p+q)}}
(\xi-\xi_0) - \frac{kt^2}{2} \right] ,
\en
which can be considered to be a weighted average of $x(\xi)$ over the
subthreshold values of $\xi$. $k$ is used to normalise the weight factor over
the $t$ integral in each of Eqs.~\ref{eq:Apos}. Also, 
\eq
 W = \phi \left( \frac{\xi_0}{\sqrt{p+q}} \right) + \int_{-\infty}^\infty
\frac{dt}{\sqrt{2\pi}} \phi\left( -\frac{\xi_0}{\sqrt{p}} -
t\sqrt{\frac{q}{p+q}} \right) \exp \left[ -\frac{t^2(2p+q)}{4(p+q)}
\right] \sqrt{p} .
\en
and $p$,$q$ are here $p_B$ and $q_B$ from Eq.~\ref{eq:pq}. 

We have to solve the eigenvalue equation 
\eq
\mathbf{A} \psi = \lambda \psi .
\en
The eigenvectors $\psi$ have the column-vector form
\eq
\psi = \left(
\{\delta z^{\alpha\beta} \} \
\right) \qquad (\alpha<\beta = 1,..,n+1)
\en
We now proceed as described in \cite{Alm+78}. There are three classes of
eigenvectors (and corresponding eigenvalues) -- those invariant under
interchange of all indices, those invariant under interchange of all but
one index, and those invariant under interchange of all but two
indices. These last describe the transverse mode, in which
we are interested. 

Let us consider fluctuations of the form
\eqa
\delta z^{\alpha\beta} &=& \Delta^{\alpha\beta} \qquad (\alpha < \beta =
1,..,n+1) \nonumber\\
\ena
with
\eqa
\Delta^{\alpha\beta} = \Delta \qquad \alpha,\beta \neq \alpha_0,\beta_0
\nonumber\\ 
\Delta^{\alpha_0\beta} = \Delta^{\alpha\beta_0} = \frac{2-n}{2}\Delta \qquad
\alpha \neq \alpha_0,\beta_0 \nonumber\\
\Delta^{\alpha_0\beta_0} = \frac{(2-n)(1-n)}{2}\Delta
\ena
ensuring orthogonality between the eigenvectors describing RS
and RSB fluctuations. As with \cite{Alm+78}, we have for $A^{(\alpha\beta),(\gamma\delta)}$ an
eigenvalue 
\eq
\lambda_A = P - 2Q + R
\en
with in this case $\frac{1}{2}(n+1)(n-2)$-fold degeneracy, and $P$, $Q$
and $R$ as described above. 

For $B^{(\alpha\beta),(\gamma\delta)}$, we consider fluctuations
\eqa
\delta {\tilde{z}}^{\alpha\beta} &=& c \Delta^{\alpha\beta} \qquad (\alpha <
\beta = 1,..,n+1) \nonumber\\
\ena
and obtain similarly the eigenvalue
\eq
\lambda_B = P' - 2Q' + R' ,
\label{eq:pqr}
\en
where
\eqa
P' &=& \frac{\displaystyle\partial^2 \mathcal{B}}{\displaystyle\partial
(i{\tilde{z}}^{\alpha\beta}) \partial (i{\tilde{z}}^{\alpha\beta})}
= \int_{-\infty}^{\infty} Dt \bigg[ \eta^2
\bigg]^2_{\eta}(\textstyle\frac{1}{2},t) \nonumber \\ 
Q' &=& \frac{\displaystyle\partial^2 \mathcal{B}}{\displaystyle\partial
(i{\tilde{z}}^{\alpha\beta}) \partial (i{\tilde{z}}^{\alpha\gamma})}
= \int_{-\infty}^{\infty} Dt \bigg[ \eta^2
\bigg]_{\eta}(\textstyle\frac{1}{3},t) \bigg[ \eta
\bigg]^2_{\eta}(\textstyle\frac{1}{3},t) 
\nonumber \\
R' &=& \frac{\displaystyle\partial^2 \mathcal{B}}{\displaystyle\partial
(i{\tilde{z}}^{\alpha\beta}) \partial (i{\tilde{z}}^{\gamma\delta})}
= \int_{-\infty}^{\infty} Dt \bigg[ \eta \bigg]^4_{\eta}(\textstyle
\frac{1}{4},t) 
, \nonumber \\
\label{eq:pqr2}
\ena
and $\bigg[ x(\eta) \bigg]_{\eta}(k,t)$, the weighted pattern average, is
defined as 
\eq
\bigg[ x(\eta) \bigg]_{\eta}(k,t) = \int_\eta d\eta P(\eta) x(\eta) \exp
\left[ -\frac{k}{2}{\tilde{z}}_{1} \eta^2 -kt\sqrt{{\tilde{z}}_{1}}
\eta \right] .
\en

\bibliographystyle{prsty}
\bibliography{srs}

\newpage

\newpage

\end{document}